\documentclass[prl,twocolumn,showpacs,preprintnumbers,amsmath,amssymb]{revtex4}

\usepackage{graphicx}
\usepackage{dcolumn}
\usepackage{bm}

\begin{document}


\title{Bose-Einstein condensation of alkaline earth atoms: $^{40}${Ca}}

\author{Sebastian Kraft }
\email{sebastian.kraft@ptb.de}
\author{Felix Vogt}
\author{Oliver Appel}
\author{Fritz Riehle}
\author{Uwe Sterr}
\affiliation{Physikalisch--Technische Bundesanstalt (PTB), Bundesallee 100, 38116 Braunschweig, Germany}

\date{\today}

\begin{abstract}
We have achieved Bose-Einstein condensation of $^{40}$Ca, the first for an alkaline earth element. The  influence of elastic and inelastic collisions associated with the large ground-state s-wave scattering length of $^{40}$Ca was measured. From these findings, an optimized loading and cooling scheme was developed that allowed us to condense about $2 \cdot 10^4$ atoms after laser cooling in a two-stage magneto-optical trap and subsequent forced evaporation in a crossed dipole trap within less than 3~s. The condensation of an alkaline earth element opens novel opportunities for precision measurements on the narrow intercombination lines as well as investigations of molecular states at the $^1$S--$^3$P asymptotes.
\end{abstract}

\pacs{03.75.Hh, 67.85.Hj}

\maketitle
The first Bose-Einstein condensate (BEC) of a dilute quantum gas in 1995 has opened completely new avenues in physics. In subsequent years this quantum degenerate state could be reached with different species (for a recent list of references see, e.\,g.,~\cite{fuk07a}). By far most research has been performed on alkali atomic and molecular BECs apart from hydrogen, metastable helium, chromium, and ytterbium. So far, no member of the alkaline earth elements could be brought to quantum degeneracy despite considerable effort \cite{kat01,gru02}. The alkaline earth elements have unique properties, e.\,g., their narrow intercombination transitions or their ground state without a magnetic moment.
Due to the non-degenerate ground state in $^{40}$Ca and in the other alkaline earth elements, the associated simpler molecule structure allows for more accurate investigations of collisions \cite{zin00,ciu04}. Moreover, the vanishing magnetic moment in the ground and excited state will allow for novel applications in atom interferometry not hampered by phase shifts due to magnetic fields. Calcium has a particularly large ground state scattering length that not only made it interesting but also difficult to realize a BEC.

Calcium (like the other alkaline earth elements) shares these properties with ytterbium which has a similar electronic structure \cite{tak03a} but has a five hundred-fold larger line width of the $^1$S$_0$-$^3$P$_1$ intercombination transition. This 370~Hz line width at 657~nm made calcium for some time the optical frequency standard with the lowest uncertainty in the visible \cite{ste04a}. Both technologies, optical frequency metrology and the generation of a BEC, can now be combined for novel applications.

In this letter we report on the preparation  of a $^{40}$Ca BEC. Starting point for our experiment is a magneto-optical trap (MOT) on the $^1$S$_0$ -- $^1$P$_1$ transition in the singlet system. The vanishing ground-state magnetic moment prevents sub-Doppler cooling of $^{40}$Ca in a MOT. To cool the atoms further we use a second MOT stage on the narrow intercombination line $^1$S$_0$ -- $^3$P$_1$ allowing for temperatures in the MOT as low as 15~$\mu$K. For efficient cooling we increase the scattering rate of this transition by quenching the upper state \cite{bin01a}. During both MOT stages an optical dipole trap is overlapped with the MOT. In order not to interfere with the narrow line cooling we choose a trapping wavelength close to a ``magic" wavelength \cite{kat01} where the ac Stark shifts of the states involved in the transition are exactly equal. This also ensures efficient loading of the atoms from the MOT into the dipole trap. As $^{40}$Ca has a large scattering length $a$ of $340\,\, a_\mathrm{0} < a < 700 \,\, a_\mathrm{0}$ (here $a_\mathrm{0} \approx $  53~pm denotes the Bohr radius) \cite{vog07} and with it a large three-body loss coefficient, a large volume single-beam optical trap is used for initial cooling. During evaporation the trap is transformed to a crossed dipole trap to maintain high trap frequencies. After turning off the MOT this allows us to cool the atoms to degeneracy within 1.6~s by forced evaporative cooling.

In the following we only give a brief summary of the experimental setup; more details are given in \cite{deg05a}. We decelerate a  beam of Ca atoms in a Zeeman slower. The beam of slowed atoms is collimated and deflected to the MOT region by a two-dimensional optical molasses. The Zeeman slower and the molasses as well as the first stage of the MOT work on the  $^1$S$_0$ --  $^1$P$_1$  transition at 423 nm. We are able to capture up to $10^9$ atoms in 1~s with an initial loading rate of $2.5 \cdot 10^9$ atoms/s at a temperature on the order of 1.5~mK.

For the second stage we use the spin forbidden intercombination line  $^1$S$_0$ --  $^3$P$_1$ with a line width of 370~Hz at 657~nm.
The light for the intercombination line is provided by a diode laser system  in master-slave configuration with a line width of about 1~Hz locked to a stable cavity \cite{sto06}. To quench the upper state we apply a laser at 453~nm, driving the transition $^3$P$_1$ -- $^1$D$_2$. As a single photon recoil shift of 11.6~kHz is larger than the effective line width of the quench-broadened transition we broaden the 657~nm laser line by adding a sinusoidal frequency modulation. The high frequency edge of the cooling spectrum is red-detuned with respect to the resonance by
about 280~kHz and the width of the modulated laser spectrum is 1.5~MHz. After 350~ms the atom number in the second MOT is up to  $10^8$ atoms. The temperature at the end of this MOT stage is typically 15~$\mu$K, resulting in a density of $3 \cdot 10^{10}$~cm$^{-3}$ and a phase space density of about $10^{-5}$.

To further increase the phase space density we load the atomic sample into an optical dipole trap by adding the trapping beam during the MOT stages. For optimal loading of the dipole trap the energy atoms gain while entering the trapping potential has to be removed from the sample. This is possible if laser cooling carries on while atoms are in the trapping region.
In order not to tune the atoms out of resonance, we operate the trapping laser near the magic wavelength (for circular polarized trap light at 983~nm $\pm$~12~nm). Here the ac-Stark shifts due to the trapping laser for the ground state and one Zeeman component of the excited state $^3$P$_1$ are equal  \cite{deg04}. In addition, it has to be assured that the transitions to the other Zeeman components are not shifted into resonance \cite{gra07}. Choosing  a high power laser with a  wavelength of 1030~nm assures that all transitions are shifted away from resonance such that the cooling laser remains red detuned from all transitions. The importance of cooling inside the dipole trap was demonstrated by a four times reduced capture rate when changing the laser to linear polarization for which the light is not close to a magic wavelength.

To characterize the losses limiting the number of atoms in the dipole trap we measure the storage time of captured atoms after turning off the 657~nm MOT. For long storage times we find an exponential decay with a lifetime of about 4~s due to collisions with background atoms at a pressure of $1\cdot10^{-9}$~mbar. In addition, we observe a loss rate depending linearly on the density when the frequency broadened red MOT light is present even though the magnetic field and the quench laser are turned off. We attribute these losses to photoassociation to the $^1$S--$^3$P asymptote and derive a two-body loss coefficient on the order of  $2\cdot 10^{-13}$~cm$^3$/s at an ensemble temperature of 20~$\mu$K and a total light intensity of 0.3~W/cm$^2$.

To provide the light for the optical dipole trap we use a Yb:YAG single frequency disk laser with an output power of up to 25~W developed at the Institut f\"ur Strahlwerkzeuge der Universit\"at Stuttgart / Germany. The laser beam is divided into two parts, each passing through an acousto-optical modulator to adjust the optical power. Both beams are focussed into the vacuum chamber to 1/e$^2$ beam waist radii of 33~$\mu$m. The first beam passes the chamber horizontally and its focus is aligned with the center of the MOT. The second beam intersects the first beam in its  waist at an angle of $54^\circ$ and is tilted by $20^\circ$ with respect to the horizontal plane \cite{note01}.

A severe limitation to cooling of atoms arises from density-dependent three-body losses, where two atoms form a bound pair while a third collision partner carries away the binding energy \cite{bur97,sta98c}. For atoms of mass $m$ and a large scattering length $a$, s-wave scattering into weakly bound dimers is predicted. The binding energy is given by $\epsilon \approx \hbar^2/(m a^2)$ leading to a scaling law for the three-body recombination rate $\alpha_\mathrm{rec} = C \hbar a^4/m$ \cite{fed96,bed00}. The coefficient $C$ is oscillating as a function of $a$ between 0 and 70. The associated three-body loss rate $L_3$ can be calculated by multiplying the scattering rate with the number of particles lost per scattering event.
Even though the scattering length for Ca is not strictly in the regime of very large $a$, this formalism gives a qualitative picture for our experiment. In the case of $^{40}$Ca, three-body collisions can lead to strong losses as the mass $m_\mathrm{Ca}$ of Ca is rather moderate while $a$  is large compared, e.\,g., to that of most alkali atoms. This holds even more as the binding energy of the weakest bound state is significantly lower than the initial trap depth. Therefore, at least at the beginning of evaporation, the third atom which carries away 2/3 of the energy is not immediately lost from the trap but distributes this energy among the trapped ensemble.

\begin{figure}[t]
\includegraphics[width=7cm]{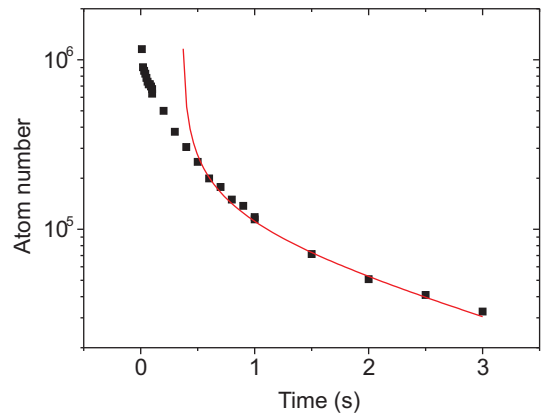}	
\caption{Atom number in the crossed dipole trap as a function of storage time (squares). The line shows a fit to the data for times $>$ 500 ms taking into account losses due to background collisions and three-body effects (see text).}
\label{fig:threebody}
\end{figure}
To measure three-body losses we load atoms directly into  a crossed dipole trap with trap frequencies of $\omega_1 = 2\pi \cdot 1.0$~kHz, $\omega_2 = 2\pi \cdot 2.0$~kHz, and $\omega_3 = 2\pi \cdot 2.2$~kHz and a trap depth of 130~$\mu$K.  We fit the decay of the atom number taking into account losses due to three-body recombination and background losses with rate $\gamma$ (Fig.~\ref{fig:threebody}). The change in the density $\rho$ is described by  $\dot{\rho} = -\gamma  \rho - L_3\rho^3$. During the first 500 ms the atoms are accumulating in the intersection of the beams. Thus, the fit was performed for storage times $t\geq500$~ms. We estimate the three-body loss coefficient to be $L_3 = 3 \cdot 10^{-27}$cm$^6$/s. As the trap is not deep enough to prevent evaporation this value is an upper bound for $L_3$.

\begin{figure}[t]
\includegraphics[width=8cm]{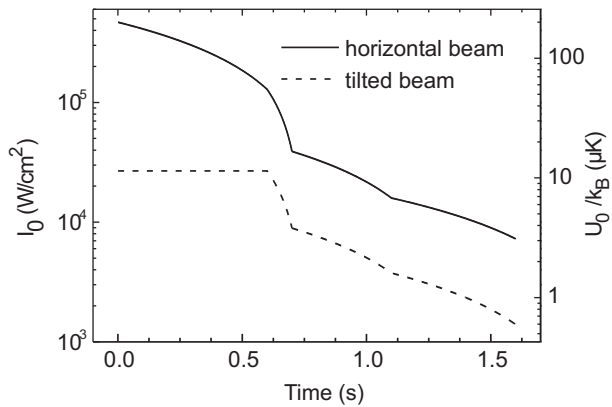}	
\caption{Ramping of the trapping beam intensities during forced evaporation. Depicted are the light intensity at the center of the atomic cloud $I_0$ and the energy of the corresponding light shift $U_0$ in units of $k_\mathrm{B}$.
In the beginning of evaporation the trap depth is determined by the light shift of the horizontal beam. At the end of ramping due to gravity the trap depth is less than $U_0$. }
\label{fig:ramp}
\end{figure}
For the generation of a BEC we load the atoms in a large volume dipole trap in order to reduce three-body losses. The power in the horizontal beam is 7~W, corresponding to a trap depth of 180~$\mu$K with radial and axial trap frequencies measured to be $\omega_r = 2 \pi \cdot 1.8$~kHz and $\omega_a = 2 \pi \cdot 7$~Hz, respectively. The focus of the tilted beam  is shifted from the intersection with the horizontal beam. In this way, the radius of the tilted beam at the position of the atoms is 77 $\mu$m. The tilted beam at an initial power of 2.5~W is not able to support atoms against gravity and therefore does not act as a trap of its own.  We are able to load $5 \cdot 10^6$ atoms into this trap at a  temperature of 20 $\mu$K after 20 ms of thermalization. The trap depth is ramped down in four linear ramps by reducing the power in the laser beams. In each ramp the trap depth is lowered by about a factor of three (Fig.~\ref{fig:ramp}).  Initially, the atoms spread out in the trap generated by the horizontal beam. During the first ramp the power of the tilted beam is  kept constant, forcing the atoms to accumulate in the crossed region during ramping. The duration of each ramp is optimized in the experiment. Within 1.6~s of forced evaporation we achieve Bose-Einstein condensation. To produce a pure BEC the intensity in the laser beams can be reduced further.

\begin{figure}[t]
\includegraphics[width=8cm]{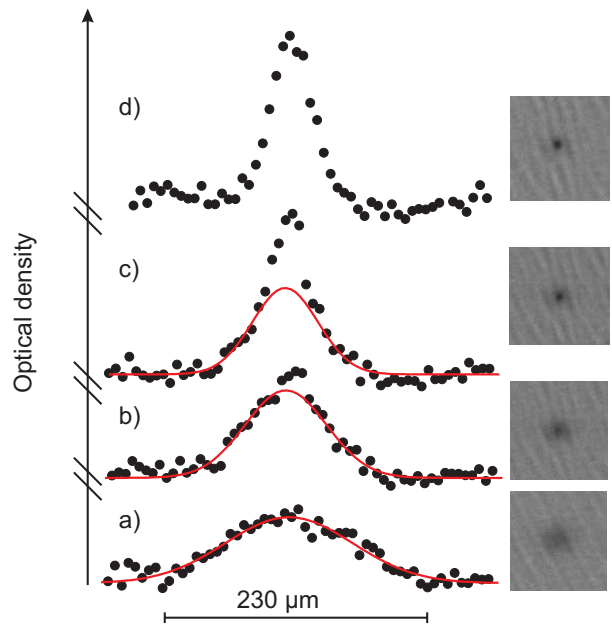}	
\caption{Density distribution integrated along the direction of the imaging beam after 7~ms of free expansion. The figure shows the density distribution along a vertical cut through the center and the corresponding absorption image. The final trap depth varies for the four figures. a) shows a thermal cloud at 260~nK. In b) the onset of Bose-Einstein condensation is visible. Cooling slightly further leads to a pronounced bimodal density distribution (c). In d) the atoms are cooled to a nearly pure BEC. }
\label{fig:profiles}
\end{figure}
The formation of the BEC becomes visible in the density distribution of the atomic cloud. Figure \ref{fig:profiles} shows vertical cuts through the center of the cloud taken after 7~ms of free expansion alongside with corresponding absorption images.  In Fig.~\ref{fig:profiles} a) the final power in the horizontal laser beam  is 100~mW, corresponding to a trap depth of 0.8~$\mu$K. The density profile as well as  the absorption image shows the typical Gaussian distribution of a thermal cloud. The ensemble consists of $9\cdot10^4$ atoms at a temperature of 260~nK. In b) the laser intensity is decreased to 90~mW corresponding to a trap depth of 0.6~$\mu$K. The cut through the density profile still shows a nearly Gaussian cloud with a small density increase in the center. The temperature of this cloud calculated by a fit to the wings of the density distribution is 170~nK. The atom number in the total cloud is $7 \cdot 10^4$; the trap frequencies are $\omega_r = 2 \pi \cdot 200$~Hz and $\omega_{a} = 2 \pi \cdot 40$~Hz. The density increase in the center of the cloud becomes clearer in c) where a bimodal distribution typical for the coexistance of a condensate and a thermal cloud is visible. The temperature measured by a fit to the Gaussian wings (solid line) is 140~nK. The trap depth is 340 nK. The total atom number in this cloud is $4\cdot 10^4$. Figure d) shows a nearly pure condensate consisting of $10^4$ atoms after further lowering of the trap depth to 150 nK.

From Fig \ref{fig:profiles} b) we can estimate the critical temperature $T_C$ where the phase transition to a BEC occurs as 170~nK in this measurement run. This is consistent with a calculation of $T_C =(\hbar\bar{\omega}/k_B)(N/1.202)^{1/3}$ = 200 nK where $\bar{\omega}$ is the geometrical average of the trap frequencies and N the atom number \cite{ket99}.

\begin{figure}[t]
\includegraphics[width=8cm]{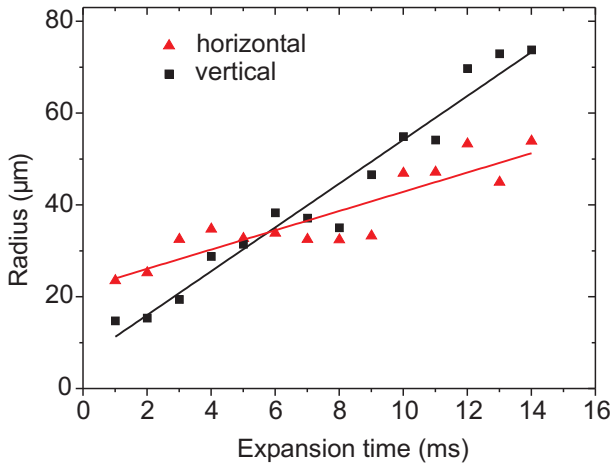}	
\caption{Free expansion of the condensate. Plotted are the Thomas-Fermi radii as a function of expansion time and linear fits through these data points. The expansion in the vertical (radial) direction is faster than the horizontal (axial) expansion due to the stronger initial confinement in the trap. }
\label{fig:expansion}
\end{figure}
Besides the bimodal density distribution, a signature of condensates is their typical anisotropic expansion \cite{cas96}. Figure \ref{fig:expansion} shows the expansion of a condensate from a dipole trap with a radial trap frequency of $\omega_r = 2\pi \cdot 190$~Hz. Due to the stronger radial confinement in the trap the expansion in vertical (radial) direction with 4.7(3) $\mu$m/ms is clearly faster than the expansion in horizontal (axial) direction with 2.0(3) $\mu$m/ms. From the temporal evolution of the expansion we can estimate the chemical potential  to be $\mu / k_B = 55$~nK, assuming that the condensate can be treated in the Thomas-Fermi limit. The calculated chemical potential satisfies $\mu \gg \hbar \omega_{r,a}$, confirming this assumption. From the chemical potential we can calculate the scattering length as $a = (2\mu)^{5/2}/(15N\hbar^2\bar{\omega}^3\sqrt{m_\mathrm{Ca}})$. From this we estimate the  scattering length of $a \approx 440~a_\mathrm{0}$  which is in good agreement with our previously measured value of 340~$a_\mathrm{0} < a  < 700$~$a_\mathrm{0}$, obtained from photoassociation measurements \cite{vog07}.

In conclusion, we have presented the first Bose-Einstein condensate of a group II element.
We have developed an optimized evaporation strategy to cope with the large three-body loss.
As a typical alkaline earth atom, calcium offers unique properties, such as highly forbidden optical transitions.
The intercombination line from the ground state to the $^3$P$_1$ level has a line width of 370~Hz and the optical transition to the $^3$P$_0$ state is strictly forbidden for single-photon excitation, but acquires a tunable transition strength by a magnetic field \cite{tai06}.
Thus this BEC offers a wealth of applications for precision measurements.
For instance, the ultra narrow transition to the $^3$P$_0$ state can be used in a magic wavelength optical lattice clock with a Mott insulator with well defined atom number per lattice site \cite{fuk09}. For unity occupation possible collisional shifts can be avoided \cite{cam09}.

In addition, the $^1$S$_0$-$^3$P$_1$ intercombination transition can be used for manipulation of the scattering length by optical Feshbach resonances \cite{ciu05,nai06}, e.\,g., to create local defects and inhomogeneities in a BEC or in a trapped lattice gas. Due to the narrow line width inelastic losses are expected to be strongly suppressed, and the creation of ultracold Ca$_2$ molecules in the vibrational ground state can occur with large probability \cite{ciu04}.
Laser cooling to the Bose-condensed state has been discussed theoretically in the ``festina-lente regime" \cite{cir96} and calcium would be an ideal candidate to test this proposal.

\begin{acknowledgments}
We thank Jun Ye for stimulating discussions and W. Ertmer and E. Rasel for loan of the disk laser. This work was financially supported by DFG through SFB 407. U.S. and F.R. are members of the Centre for Quantum Engineering and Space-Time Research QUEST.

\end{acknowledgments}

\begin{thebibliography}{10}
\bibitem{note01}
The particular choice of the tilting angle is due to limited optical access of the vacuum chamber.
\end{thebibliography}

\end{document}